\documentclass[prd,twocolumn,nofootinbib,nobibnotes,noeprint,preprintnumbers,floatfix]{revtex4-2}
\usepackage{amssymb,amsmath,graphicx}
\usepackage[dvipsnames]{xcolor}
\usepackage[colorlinks,allcolors=NavyBlue]{hyperref}
\usepackage{microtype}

\newcommand{\angl}[1]{ \left\langle #1 \right\rangle }
\newcommand{\Neff}{ N_\mathrm{eff} }
\newcommand{\DNeff}{ \Delta N_\mathrm{eff} }

\begin{document}

\preprint{FERMILAB-PUB-22-913-T}

\title{Twin Sterile Neutrino Dark Matter}

\author{Ian Holst$^{1,2}$}
\thanks{holst@uchicago.edu, https://orcid.org/0000-0003-4256-3680}

\author{Dan Hooper$^{1,2,3}$}
\thanks{dhooper@fnal.gov, http://orcid.org/0000-0001-8837-4127}

\author{Gordan Krnjaic$^{1,2,3}$}
\thanks{krnjaicg@fnal.gov, http://orcid.org/0000-0001-7420-9577}

\author{Deheng Song$^{3,4,5}$}
\thanks{songdeheng@yukawa.kyoto-u.ac.jp, 0000-0003-3441-4212}

\affiliation{$^1$Department of Astronomy and Astrophysics, University of Chicago, Chicago, Illinois 60637, USA}
\affiliation{$^2$Kavli Institute for Cosmological Physics, University of Chicago, Chicago, Illinois 60637, USA}
\affiliation{$^3$Theoretical Astrophysics Group, Fermi National Accelerator Laboratory, Batavia, Illinois 60510, USA}
\affiliation{$^4$Center for Neutrino Physics, Department of Physics, Virginia Tech, Blacksburg, Virginia 24061, USA}
\affiliation{$^5$Center for Gravitational Physics and Quantum Information, Yukawa Institute for Theoretical Physics, Kyoto University, Kyoto, Kyoto 606-8502, Japan}

\date{\today}

\begin{abstract}
We propose that the dark matter of our Universe could be sterile neutrinos which reside within the twin sector of a mirror twin Higgs model. In our scenario, these particles are produced through a version of the Dodelson-Widrow mechanism that takes place entirely within the twin sector, yielding a dark matter candidate that is consistent with X-ray and gamma-ray line constraints. Furthermore, this scenario can naturally avoid the cosmological problems that are typically encountered in mirror twin Higgs models. In particular, if the sterile neutrinos in the Standard Model sector decay out of equilibrium, they can heat the Standard Model bath and reduce the contributions of the twin particles to $\Neff$. Such decays also reduce the effective temperature of the dark matter, thereby relaxing constraints from large-scale structure. The sterile neutrinos included in this model are compatible with the seesaw mechanism for generating Standard Model neutrino masses.
\end{abstract}

\maketitle

\section{Introduction}

Despite its considerable empirical successes, the Standard Model (SM) does not explain a number of observed phenomena, including the origin and identity of dark matter (DM), the existence of neutrino masses, and the large hierarchy between the electroweak and Planck scales. In the SM, radiative corrections are naively expected to increase the mass of the Higgs boson to near the Planck scale, leading to what is known as the electroweak hierarchy problem.

Supersymmetry, if broken near the TeV scale, could facilitate the cancellation of the largest of these corrections, but this class of solutions to the hierarchy problem appears increasingly fine-tuned in light of null results from the Large Hadron Collider~\cite{ATLAS:2017drc,ATLAS:2017eoo,ATLAS:2018qzb,CMS:2018xqw,CMS:2017jrd,CMS:2017kil}. An alternative approach, known as ``neutral naturalness,'' stabilizes the electroweak hierarchy by introducing new symmetries and new particles without color or other SM gauge charges, making it possible to cancel the quadratic contributions to the Higgs mass while simultaneously satisfying current constraints from colliders~\cite{CraigNeutralNaturalnessOrbifold2015,Barbieri:2005ri,Chacko:2005vw,Burdman:2006tz,Cai:2008au,Poland:2008ev,Craig:2015pha,Cohen:2018mgv,Cheng:2018gvu,BatellReviewNeutralNaturalness2022}.

The prototypical example of neutral naturalness is the twin Higgs framework~\cite{ChackoNaturalElectroweakBreaking2006,Barbieri:2005ri,Chacko:2005vw,Craig:2015pha,Chacko:2005un,Perelstein:2005ka,Craig:2013fga,CraigNeutralNaturalnessOrbifold2015}, which introduces ``twin'' copies of SM particles that are charged under twin versions of the SM gauge group. A discrete ``twin symmetry'' between the SM and twin sector parameters helps to stabilize the Higgs mass up to one-loop order, allowing the hierarchy problem to be addressed at a higher energy scale, such as through a supersymmetric UV completion~\cite{Falkowski:2006qq,Chang:2006ra,Craig:2013fga}. For such a model to be phenomenologically viable, the twin symmetry must be softly broken, allowing the Higgs vacuum expectation value (VEV) to be different in each sector. For a ratio of twin-to-SM VEVs of $v_B/v_A \sim 3-5$, such a scenario can address the hierarchy problem while allowing the 125 GeV Higgs boson to behave approximately as predicted by the SM, containing only a $\sim v_A^2/v_B^2 \sim \mathcal{O}(10\%)$ admixture of the twin sector Higgs boson.

Even with asymmetric VEVs, cosmological problems can arise in ``mirror twin Higgs'' (MTH) models, in which the SM and twin sectors contain the same particle content. Because the twin and SM sectors are initially in equilibrium, large energy densities of twin photons and twin neutrinos will be present in the early Universe in MTH models, increasing the Hubble expansion rate in conflict with measurements of the effective number of neutrino species, $\Neff$. Proposed solutions to this problem include introducing asymmetric and out-of-equilibrium decays of new particles to dilute the energy density in the twin sector~\cite{ChackoCosmologyMirrorTwin2017,Craig:2016lyx}, reducing the number of twin species as in ``fraternal twin Higgs'' models~\cite{Craig:2015pha,GarciaGarcia:2015fol}, or delaying the thermal decoupling of the two sectors through SM-twin neutrino mixing \cite{CsakiViableTwinCosmology2017}.

A variety of DM candidates can arise within the context of twin Higgs models~\cite{Craig:2015xla,GarciaGarcia:2015fol,Curtin:2021spx,Farina:2015uea,Farina:2016ndq,Freytsis:2016dgf,GarciaGarcia:2015pnn,Badziak:2019zys,Cheng:2018vaj,Hochberg:2018vdo,Beauchesne:2020mih,Koren:2019iuv,Terning:2019hgj,Curtin:2021alk,Ritter:2021hgu,Ahmed:2020hiw,Badziak:2022eag,Kilic:2021zqu}, including the twin tau, which in fraternal models can be stable, functioning as a prototypical WIMP~\cite{Craig:2015xla,GarciaGarcia:2015fol,Curtin:2021spx}. However, in light of increasingly stringent constraints from LZ and other direct detection experiments~\cite{LZ:2022ufs}, it is well-motivated to consider other dark matter candidates that might arise in the twin Higgs framework.

The problem of non-zero neutrino masses can be simply explained by supplementing the SM with right-handed neutrinos. Through the seesaw mechanism~\cite{MinkowskiSeesaw1977,Yanagida:1979as,Mohapatra:1979ia}, this can lead to light ``active'' neutrinos which are mostly left-handed, and heavier ``sterile'' states which are mostly right-handed. Being sufficiently long-lived, cold, and feebly interacting, sterile neutrinos have long been seen as an attractive candidate for DM~\cite{DodelsonSterileNeutrinosDark1994,Shi:1998km,Dolgov:2000ew,AbazajianSterileNeutrinoHot2001,BoyarskySterileNeutrinoDark2019}. They can be generated in the early Universe through the Dodelson-Widrow mechanism, in which active neutrinos oscillate, interact, and collapse into a sterile state~\cite{DodelsonSterileNeutrinosDark1994}.

The simplest Dodelson-Widrow scenario has been entirely ruled out in recent years by searches for DM decaying to X-ray~\cite{Roach:2022lgo,Foster:2021ngm,Roach:2019ctw,Ng:2019gch,Perez:2016tcq,Neronov:2016wdd,Boyarsky:2007ge,Yuksel:2007xh} or gamma-ray lines~\cite{Fermi-LAT:2015kyq}, as well as by constraints on warm DM from observations of small-scale structure~\cite{NadlerConstraintsDarkMatter2021,Dekker:2021scf,HoriuchiSterileNeutrinoDark2014,Schneider:2016uqi}. For sterile neutrinos to be a viable DM candidate, they must be produced through resonant~\cite{Shi:1998km} or non-thermal~\cite{Kusenko:2006rh,Merle:2013wta,Merle:2015oja,Shaposhnikov:2006xi,Petraki:2007gq,Abada:2014zra} mechanisms, or involve other exotic elements ~\cite{Berlin:2016bdv,deGouveaDodelsonWidrowMechanismPresence2020,BringmannMinimalSterileNeutrino2023}.

In this paper, we propose incorporating sterile neutrinos into a mirror twin Higgs model, identifying the twin sterile neutrino as a new DM candidate. We show that by combining these two frameworks, we can address the hierarchy problem, explain the origin of neutrinos masses, and provide a viable DM candidate, while resolving each of these frameworks' respective tensions with cosmological and astrophysical observations. In this scenario, the twin sterile neutrino DM is produced through a Dodelson-Widrow mechanism taking place entirely within the twin sector. When the DM decays, it produces few observable SM photons, evading X-ray and gamma-ray constraints. Meanwhile, a heavier SM sector sterile neutrino decays and heats the SM bath~\cite{ChackoCosmologyMirrorTwin2017}, reducing the relative energy density in the twin sector and, thus, solving the $\Neff$ problem of the MTH model.

\section{Model}
We consider a minimal MTH model whose twin-symmetric Lagrangian is supplemented with the terms
\begin{equation*} \label{lagrangian}
    \mathcal{L} \supset - \!\!\! \sum_{i=A,B}\! \left( y_\nu^i H^i l_L^i \nu_R^i + \frac{M_i}{2} \nu_R^i \nu_R^i \right) \!-\! M_{AB}\, \nu_R^A \nu_R^B + \mathrm{h.c.}~,
\end{equation*}
where $A$ and $B$ represent the SM and twin sectors, respectively. For each sector, $H^i$ is the Higgs doublet, $l_L^i$ is the lepton doublet that contains a left-handed neutrino $\nu^i_{L}$, $\nu_R^i$ is the right-handed neutrino, and $y_\nu^i$ is the Yukawa coupling. We allow for explicit twin symmetry breaking in the neutrino sector, such that $M_A \ne M_B$ and $y^A_{\nu} \ne y^B_{\nu}$. The last term allows the right-handed neutrinos $\nu_R^A$ and $\nu_R^B$ to mix between sectors through the coefficient $M_{AB}$. For simplicity, we consider only one generation of neutrinos, although this scenario can be extended to the case of three generations.

After electroweak symmetry breaking, the Yukawa terms yield Dirac masses of $m_i = y_\nu^i v_i / \sqrt{2}$. We adopt $v_B/v_A = 3$ throughout, where the SM Higgs VEV $v_A = 246$~GeV. We also assume a mass hierarchy of $M_A, M_B \gg M_{AB} \gg m_A, m_B$, so that the $4 \times 4$ neutrino mass matrix is diagonalized via an active-sterile seesaw mechanism~\cite{MinkowskiSeesaw1977,Yanagida:1979as,Mohapatra:1979ia} within each sector, as well as a SM-twin seesaw which mixes the sterile states between the two sectors. To leading order in the small mixing angles, this yields two active neutrino mass eigenstates,
\begin{equation}
    \begin{split}
        \nu_A &= \nu_L^A - \theta_A\,\nu_R^A\\
        \nu_B &= \nu_L^B - \theta_B\,\nu_R^B,
    \end{split}
\end{equation}
and two sterile neutrino mass eigenstates,
\begin{equation}
    \begin{split}
        N_A &= \nu_R^A + \theta_A\,\nu_L^A + \theta_{AB}\,\nu_R^B\\
        N_B &= \nu_R^B + \theta_B\,\nu_L^B - \theta_{AB}\,\nu_R^A~.
    \end{split}
\end{equation}
The mixing angles can be written
\begin{equation}
    \theta_A = \frac{m_A}{M_A},~~~ \theta_B = \frac{m_B}{M_B},~~~ \theta_{AB} = \frac{M_{AB}}{|M_A - M_B|},
\end{equation}
and the mass eigenvalues are
\begin{equation}
    \begin{split}
        m_{\nu_A} &= \frac{m_A^2}{M_A}, ~~~~~~ M_{N_A} = M_A,\\
        m_{\nu_B} &= \frac{m_B^2}{M_B}, ~~~~~~ M_{N_B} = M_B,
    \end{split}
\end{equation}
demonstrating the familiar seesaw relation in each sector. We identify the twin sterile neutrino $N_B$ as the dark matter candidate and the SM sterile neutrino $N_A$ as a heavier state that decays in the early Universe.

Note that our model contains the same field content as the model in Ref.~\cite{ChackoCosmologyMirrorTwin2017}, which preserved a full twin symmetry between the two sectors and did not consider DM production. However, for sterile neutrino DM in a twin Higgs context, an exact twin symmetry in the neutrino parameters combined with $M_{AB}$ mixing is problematic because it maximally mixes the sterile states between the sectors ($M_A = M_B \implies \theta_{AB} = \pi/4$). Since the Fermi constant $G_{F,i} = v_i^{-2} / \sqrt{2}$ is larger in the SM sector, the maximally mixed DM decays predominantly to SM particles, including monoenergetic photons. Such a model is sufficiently similar to the canonical Dodelson-Widrow scenario that it yields comparable X-ray and gamma-ray line signals, which are already excluded by observations.

In contrast, our model overcomes this limitation by explicitly breaking the twin symmetry between the SM and twin neutrino parameters. In the $M_{AB} \ll M_A, M_B$ limit, having $M_A \gg M_B$ not only ensures that $\theta_{AB}$ is small, but also that the lighter sterile neutrino $N_B$ (our DM candidate) couples mostly to the twin sector and is long-lived, while the heavier sterile state $N_A$ couples mostly to the SM and decays in the early Universe.

\begin{figure*}
    \centering
    \includegraphics[width=\linewidth]{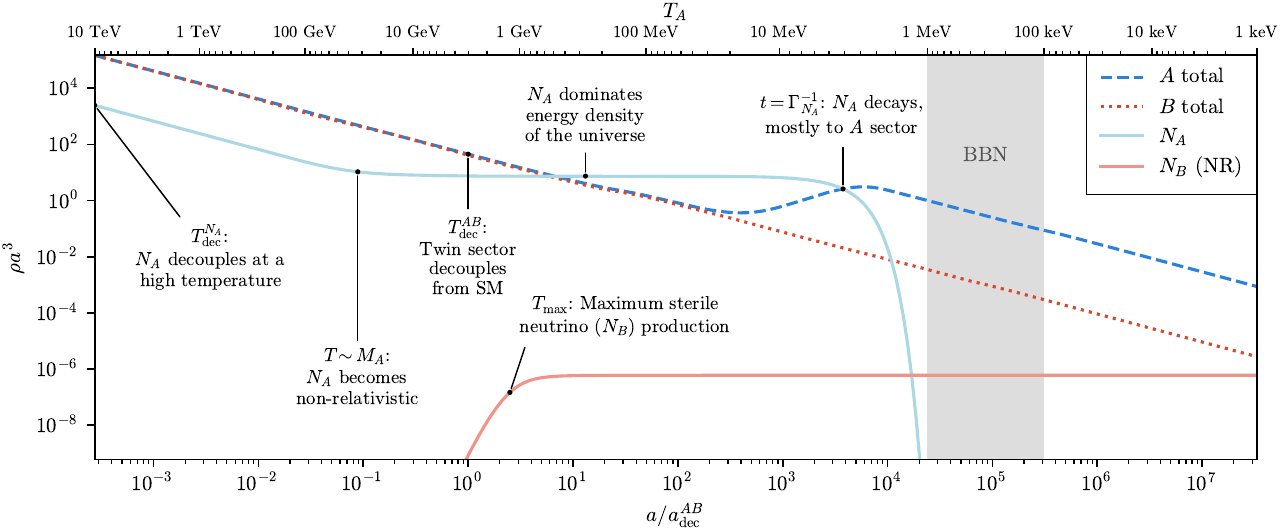}
    \caption{Timeline of events in one realization of our scenario. The curves indicate the comoving energy densities of radiation in the SM ($A$) and twin ($B$) sectors, and of the sterile neutrinos, $N_A$ and $N_B$, as a function of scale factor, $a$, or the temperature of the SM bath, $T_A$. The $y$ axis is normalized by the comoving energy density of the SM sector at $T_A = 1~\mathrm{MeV}$. Only the non-relativistic portion of $\rho_{N_B} = M_{N_B} n_{N_B}$ is shown. This example takes the parameters $M_{N_A} = 100~\mathrm{GeV}$, $m_{\nu_A} = 10^{-9}~\mathrm{eV}$, $M_{N_B} = 100~\mathrm{keV}$, and $m_{\nu_B} = 4 \times 10^{-4}~\mathrm{eV}$, for which we find $\DNeff \approx 0.02$ and $\Omega_{N_B} \approx 0.26$, consistent with observations.}
    \label{timeline}
\end{figure*}

\pagebreak

\section{Cosmology}

The cosmic history in our scenario is schematically illustrated in Fig.~\ref{timeline} and proceeds as follows:
\begin{enumerate}
    \item The $N_A$ sterile neutrino population is initially in equilibrium with the thermal bath of SM and twin particles, but it eventually decouples at temperature $T^{N_A}_{\rm dec}$.

    \item At $T \sim M_A$, the $N_A$ population becomes non-relativistic and its fraction of the total energy density begins to grow as the Universe cools and expands.

    \item Higgs portal interactions, due to $H^A$-$H^B$ mixing, keep the two sectors in equilibrium until $T_\mathrm{dec}^{AB}~\approx 3~\mathrm{GeV}$~\cite{ChackoCosmologyMirrorTwin2017}, when the SM and twin sectors decouple.

    \item After $AB$ decoupling, the $N_A$ population eventually dominates the energy density of the Universe.

    \item The $N_A$ population decays out of equilibrium to heat the SM radiation bath. This entropy injection increases the SM temperature and effectively dilutes the relative energy density of the twin sector. Note that steps 1--5 must occur prior to Big Bang nucleosynthesis (BBN) to ensure that any contributions to $\Neff$ are minimal.

    \item At some time (around $T_{\rm max}$ in Fig.~\ref{timeline}) before the epoch of matter-radiation equality, twin sterile neutrino dark matter $N_B$ is produced through active-sterile neutrino oscillations (\textit{i.e.}~a Dodelson-Widrow mechanism) within the twin sector.
\end{enumerate}

The temperature evolution in the two sectors and the energy density in $N_A$ is described by the following system of differential equations (see Appendix A for a derivation):
\begin{align}
    \frac{d\ln T_A}{d\ln a} &= -\frac{1 - \cfrac{\angl{\Gamma_{N_A \to A}} \rho_{N_A}}{3H s_A T_A} }{1 + \cfrac{1}{3}\cfrac{d\ln g_{\star,s}^{A}}{d\ln T_A}} \label{temperature_equation_A}\\
    \frac{d\ln T_B}{d\ln a} &= -\frac{1 - \cfrac{\angl{\Gamma_{N_A \to B}} \rho_{N_A}}{3H s_B T_B} }{1 + \cfrac{1}{3}\cfrac{d\ln g_{\star,s}^{B}}{d\ln T_B}} \label{temperature_equation_B}\\
    \frac{d\rho_{N_A}}{dt} &= -3H(\rho_{N_A} + P_{N_A}) - \angl{\Gamma_{N_A}} \rho_{N_A}, \label{NA_equation}
\end{align}
where $a$ is the scale factor, $T_i$ is the temperature of the radiation bath in sector $i$, $H =(8\pi G \rho/3)^{1/2}$ is the Hubble rate, $\rho = \rho_A + \rho_B$ is the total energy density, $\rho_i$ is the energy density in either sector, $s_i$ is the entropy density in either sector, and $P_{N_A}$ is the pressure of the $N_A$ population. $\angl{\Gamma_{N_A \to i}}$ is the thermally-averaged decay width of $N_A$ to sector $i$ (including the effects of time dilation). At rest, the $N_A$ decay width to SM particles is
\begin{equation} \label{gamma_NA}
    \Gamma_{N_A \to A} = C_A \frac{G_{F,A}^2}{192\pi^3} \theta_A^2 M_{N_A}^5,
\end{equation}
where $C_A$ is a factor that is obtained by summing over the decay channels (see Appendix A of Ref.~\cite{ChackoCosmologyMirrorTwin2017}). This depends on the full multi-generation neutrino mass matrix, but for one generation, we find $C_A \approx 12.4$ for $M_{N_A} = 100$ GeV. For all cases we consider here, $\Gamma_{N_A \to B} \ll \Gamma_{N_A \to A}$, so we neglect entropy transfer from $N_A$ to the twin sector. Note that the fermionic contributions to the effective relativistic degrees of freedom in energy and entropy, $g_{\star}$ and $g_{\star, s}$, are different in each sector due to the larger masses of the twin fermions, while the QCD phase transition takes place at approximately the same temperature in both sectors~\cite{BorsanyiLatticeQCDCosmology2016}. We numerically solve Eqs.~(\ref{temperature_equation_A})--(\ref{NA_equation}) to find $T_A(a)$, $T_B(a)$, and $\rho_{N_A}(a)$.

The main constraints on this scenario arise from measurements of the energy density in dark radiation, parametrized in terms of the contribution to the effective number of relativistic neutrino species, $\DNeff$, which we define as
\begin{equation} \label{Neff}
    \DNeff = \frac{\rho_{B} + \rho_{N_A}}{\rho_\nu},
\end{equation}
where $\rho_\nu$ is the energy density of a single SM active neutrino species. Note that this contribution to $\Neff$ is not from the energy injected by the decays of $N_A$, but rather from the energy density of the twin sector particles that remain after the $N_A$ population has decayed (as well as any $N_A$ particles that have not yet decayed).

\begin{figure}
    \centering
    \includegraphics[width=\columnwidth]{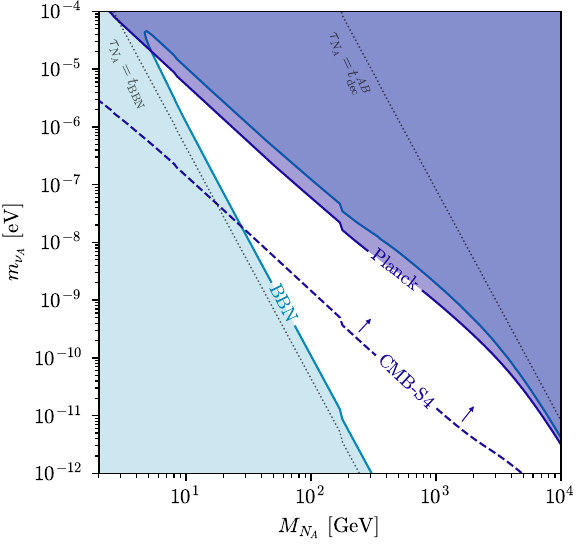}
    \caption{Constraints on the masses of the decaying sterile neutrino, $N_A$, and the corresponding active neutrino, $\nu_A$, from measurements of $\Neff$. The shaded regions represent the parameter space that is ruled out by measurements of the light element abundances~\cite{Cyburt:2015mya} (light blue) and the cosmic microwave background (dark blue). Current $\DNeff^\mathrm{CMB}$ constraints are shown from Planck~\cite{Planck2018} as well as the projected sensitivity of Stage 4 CMB experiments~\cite{CMB-S4:2016ple}. In the unshaded regions, the decays of the $N_A$ dilute the energy density of the twin sector to acceptable levels. Dotted gray lines represent curves of constant $\tau_{N_A}$.}
    \label{NAconstraints}
\end{figure}

In Fig.~\ref{NAconstraints}, we plot the parameter space in this scenario. Contours are shown for values of $\DNeff$ at the time of BBN ($T_A \approx 1$~MeV) \cite{KawasakiCosmologicalConstraintsLatetime1999,KawasakiMeVscaleReheatingTemperature2000,HasegawaMeVscaleReheatingTemperature2019}, and at the time of recombination ($T_A \approx 0.3$~eV). Current 95\% confidence level constraints require $\DNeff^\mathrm{CMB} \lesssim 0.3$~\cite{Planck2018} and $\DNeff^\mathrm{BBN} \lesssim 0.4$~\cite{Cyburt:2015mya}, while Stage 4 CMB experiments are projected to be sensitive at $1\sigma$ down to $\DNeff^\mathrm{CMB} \lesssim 0.03$~\cite{CMB-S4:2016ple}. The constraints can be understood in terms of two criteria: In the upper right region of this figure, the $N_A$ population decays before the SM and twin sectors decouple, so that both sectors quickly thermalize again, leading to unacceptably large contributions to $\Neff$. In the lower left portion of the figure, $N_A$ decays during or after BBN, again leading to large $\DNeff^\mathrm{BBN}$ (but not necessarily large $\DNeff^\mathrm{CMB}$). Note that current constraints require $y_\nu^A \lesssim 10^{-9}$.

Because the rates for sterile neutrino production and decay are set by the same model parameters, there is an inherent tension between establishing a thermalized population of $N_A$ particles in equilibrium abundance (required for step 1) and having these states decay at the right time (required to minimize contributions to $\Neff$). Neither active-sterile oscillations nor Higgs-mediated Yukawa interactions can produce $N_A$ in equilibrium abundance in the parameter space of interest. To illustrate this tension, consider that thermalizing $N_A$ through the oscillation channel requires the oscillation production rate to exceed the rate of Hubble expansion:
\begin{equation}
    \Gamma_{\nu_A \to N_A} \sim \sin^2{2\theta_A^m} G_F^2 T^5 \gtrsim \frac{T^2}{M_{\rm Pl}} \sim H.
\end{equation}
Because of matter effects, $\Gamma_{\nu_A \to N_A}$ is maximized at a temperature of $T^{N_A}_\mathrm{max} \approx 0.72\, M_W^{2/3} \,M_{N_A}^{1/3}$ \cite{DodelsonSterileNeutrinosDark1994}. Requiring $\Gamma_{\nu_A \to N_A}> H$ at $T_\mathrm{max}$ and using the seesaw relation for the active neutrino mass results in the condition
\begin{equation}
    m_{\nu_A} \gtrsim \frac{ 540 M_W^2 }{M_\mathrm{Pl}} \approx 3\times 10^{-4} \rm eV ,
\end{equation}
which is excluded by measurements of $\DNeff$, as shown in Fig.~\ref{NAconstraints}. A similar argument can be made for Higgs-mediated scattering channels for $N_A$ production in the unbroken electroweak phase (e.g., $\nu_A t \to N_A t$ through $t$-channel Higgs exchange). Thus, some \textit{additional} twin symmetry breaking interaction is necessary to establish the $N_A$ population in equilibrium at early times; no such requirements apply for $N_B$ whose abundance is generated through sub-Hubble active-sterile oscillations in the twin sector. This tension and requirement for new physics also applies to the scenario outlined in \cite{ChackoCosmologyMirrorTwin2017}.

Throughout our analysis, we assume that $N_A$ freezes out from these unspecified new interactions at a sufficiently high temperature and that $N_B$ experiences no such interactions. Since we have explicitly introduced explicit twin symmetry breaking for neutrino parameters in this model, it is straightforward to add an additional new state that couples to SM sector neutrinos to bring $N_A$ into equilibrium at early times and whose impact on later cosmological epochs is negligible (e.g., because it decays long before $N_A$ decouples). Furthermore, since these additional states introduce twin symmetry breaking only in the neutrino sector, they have a negligible impact on the other desirable features of twin Higgs models (e.g., cancellation of the quadratically divergent top loop contribution to the Higgs mass).

The lower right boundary of the viable parameter space shown in Fig.~\ref{NAconstraints} depends on the specific model of these additional interactions, so we leave it unconstrained. Given a concrete realization of such interactions, a lower limit will appear, but its precise location is model-dependent. For example, this is relevant if $N_A$ freezes out at a temperature lower than its mass, since it will be exponentially suppressed as a cold relic and its decays will have limited impact on the $\DNeff$ problem.

\section{Twin Sterile Neutrino Dark Matter}

The production of twin sterile neutrino DM through the Dodelson-Widrow mechanism in the twin sector is described by the following Boltzmann equation:
\begin{equation} \label{DW_equation}
    \!\! \frac{d {n}_{N_{\!B}}}{dt} + 3Hn_{N_{\!B}} = 2 \! \int \!\! \frac{d^3p}{(2\pi)^3} \frac{\Gamma_{\nu_B}^\mathrm{int}}{4} \!\sin^2{ \! 2\theta_{\!B}^m} (f_{\nu_B} \! - \! f_{N_{\!B} \!}),
\end{equation}
where $n$ is the number density of a given species and $f$ is its phase space distribution. The quantity $\Gamma_{\nu_B}^\mathrm{int} \propto G_{F,B}^2 T_B^5$ is the scattering rate of the twin active neutrinos, which we calculate using the values tabulated in Ref.~\cite{VenumadhavSterileNeutrinoDark2016} (for the case of neutrinos of muon flavor), rescaled by a factor of $(v_B/v_A)^{-4}$. The quantity $\theta_B^m$ is the effective active-sterile mixing angle in the twin sector in the presence of matter. This is related to the corresponding mixing angle in vacuum according to the following expression~\cite{DodelsonSterileNeutrinosDark1994,AbazajianSterileNeutrinoHot2001}:
\begin{equation}
    \sin^2{2\theta_B^m} =\frac{\Delta^2 \sin^2{2\theta_B}}{\Delta^2 \sin^2{2\theta_B} + (\Delta \cos{2\theta_B} - V_T)^2},
\end{equation}
where $\Delta = (M_{N_B}^2 - m_{\nu_B}^2)/2p$ is the vacuum oscillation factor for a neutrino of momentum $p$, and the thermal potential is given by \cite{NotzoldNeutrinoDispersionFinite1988,AbazajianSterileNeutrinoHot2001}
\begin{equation}
    V_T (p) = - \frac{8\sqrt{2}\,G_\mathrm{F,B}\,p}{3} \left(\frac{\rho_{\nu_B} + \rho_{\bar{\nu}_B}}{m_{Z_B}^2} + \frac{\rho_{l_B} + \rho_{\bar{l}_B}}{m_{W_B}^2}\right),
\end{equation}
where $\rho_{\nu}$ and $\rho_{\bar{\nu}}$ denote the energy densities of the twin active neutrinos and antineutrinos, and $\rho_{l}$ and $\rho_{\bar{l}}$ are the energy densities of the twin charged leptons and antileptons of the same flavor.

In the early Universe, these matter effects make $\sin^2{2\theta_B^m}$ small, suppressing the oscillation rate of $\nu_B$ into $N_B$. The rate of $N_B$ production increases as the temperature falls, but not necessarily enough for the $N_B$ abundance to reach its equilibrium value. This continues until the expansion of the Universe begins to efficiently suppress the rate of $N_B$ production at later times, and their abundance freezes in. Most twin sterile neutrinos are produced near a temperature of $T_\mathrm{max} \approx 270\,\mathrm{MeV}\,(M_{N_B} / \mathrm{keV})^{1/3}$ (for $v_B/v_A=3$)~\cite{DodelsonSterileNeutrinosDark1994}.

For given values of $M_{N_B}$ and $m_{\nu_B}$, we numerically solve Eq.~(\ref{DW_equation}), in conjunction with Eqs.~(\ref{temperature_equation_A})--(\ref{NA_equation}), for a large number of points within the viable region in Fig.~\ref{NAconstraints}. The decays of $N_A$ must be accounted for here, because they dilute the twin sterile neutrino DM abundance along with the rest of the twin sector particles.

\begin{figure}
    \centering
    \includegraphics[width=0.99\columnwidth]{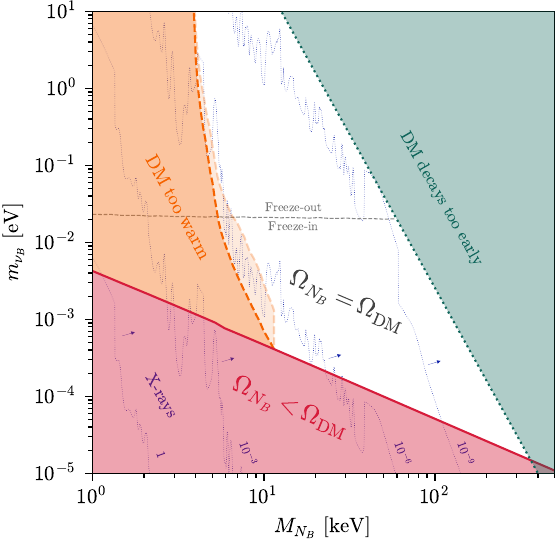}
    \caption{Constraints on the masses of our dark matter candidate, $N_B$, and the twin sector active neutrino, $\nu_B$. Shaded regions indicate constraints on the DM abundance (red), temperature (orange), and lifetime (green). Throughout the white region labeled ``$\Omega_{N_B}=\Omega_\mathrm{DM}$'' our model can produce the observed DM abundance while remaining consistent with the constraints on $\DNeff$. The dotted blue lines represent constraints on DM decay to X-ray lines for indicated values of the branching ratio $\Gamma_{N_B \to \nu_A \gamma_A} / \Gamma_{N_B \to \nu_B \gamma_B}$; for each choice of this ratio, the parameter space to the right is excluded.}
    \label{NBconstraints}
\end{figure}

In Fig.~\ref{NBconstraints}, we summarize the constraints on twin sterile neutrino DM. Throughout the white region labeled as ``$\Omega_{N_B}=\Omega_\mathrm{DM}$'', there exists some combination of $M_{N_A}$ and $m_{\nu_A}$ that lead to the observed DM abundance ($\Omega_\mathrm{DM} \approx 0.26$~\cite{Planck2018}) through the twin sector Dodelson-Widrow mechanism while remaining consistent with the constraints shown in Fig.~\ref{NAconstraints}. We do not consider resonant production mechanisms that rely on significant lepton number asymmetry~\cite{Shi:1998km}, although these could further expand this region. In the lower red region, the DM is consistently underproduced for all viable values of $M_{N_A}$ and $m_{\nu_A}$.

Another important constraint on sterile neutrino DM arises from its impact on large-scale structure formation. In the standard Dodelson-Widrow scenario, sterile neutrinos with masses smaller than 25 keV~\cite{NadlerConstraintsDarkMatter2021} are unacceptably warm, leading to fewer satellite galaxies than are observed~\cite{HoriuchiSterileNeutrinoDark2014,Schneider:2016uqi,CherryClosingResonantlyProduced2017}. In our scenario, however, the decays of $N_A$ reduce $T_B/T_A$, effectively cooling the DM and relaxing this constraint to $M_{N_B, \mathrm{min}} = 25\,\mathrm{keV}\,(T_B/T_A )$. Throughout the orange shaded region in Fig.~\ref{NBconstraints}, the DM is too warm for all viable combinations of $M_{N_A}$ and $m_{\nu_A}$, while the light orange region is too warm for only some of those parameter values. In the green region, the DM decays too rapidly, with $\tau_{N_B} < 288$ Gyr~\cite{BuckoConstrainingDarkMatter2023}, in conflict with observations of large-scale structure~\cite{BuckoConstrainingDarkMatter2023,SimonConstrainingDecayingDark2022,Poulin:2016nat,AudrenStrongestModelindependentBound2014}.

So far, we have neglected the effects of mixing between the sectors, but if $\theta_{AB} \neq 0$, constraints can arise from the null results of X-ray line searches~\cite{BoyarskyStrategySearchingDark2006,Yuksel:2007xh,HoriuchiSterileNeutrinoDark2014,Roach:2019ctw,Foster:2021ngm,Roach:2022lgo,CaloreConstraintsLightDecaying2023}, which are sensitive to loop-induced decays of $N_B$ to an active neutrino and a monoenergetic SM photon~\cite{PalRadiativeDecaysMassive1982,AdhikariWhitePaperKeV2017,BoyarskySterileNeutrinoDark2019}. These constraints are shown in Fig.~\ref{NBconstraints} as dotted blue lines. If the branching fraction for $N_B \to \nu_A \gamma_A$ (which is related to $\theta_{AB}$) is small, this signal can be highly suppressed, well below the reach of existing telescopes.

In some regions of parameter space, $\sin^2{2\theta_B^m}$ is large enough that an $N_B$ population can be produced with an equilibrium abundance, so instead of freezing in, it freezes out as a hot relic. The approximate boundary between these two cases is indicated in Fig.~\ref{NBconstraints}, showing that the observed DM abundance can be obtained in both cases.

\section{Summary}

In this paper, we have proposed a new dark matter candidate in the form of a sterile neutrino in the twin sector of a mirror twin Higgs model. This dark matter candidate can be generated through neutrino oscillations in the twin sector. If the mixing between the Standard Model and twin sectors is sufficiently small, the decays of the dark matter particles will not produce an observable X-ray or gamma-ray line. Furthermore, a sterile neutrino in the Standard Model sector can naturally undergo out-of-equilibrium decays in this model, diluting the energy density of the twin sector, and thus evading constraints from $\Neff$. We emphasize that measurements of $\Neff$ still constrain the neutrino mass parameters such that new interactions are needed to establish the SM sterile neutrino population in equilibrium. These decays can also relax the constraints on warm dark matter that arise from observations of large-scale structure. This scenario can restore both mirror twin Higgs models and sterile neutrino DM as viable solutions to several long-standing mysteries in particle physics and cosmology, while remaining consistent with all existing constraints.

\bigskip

\begin{acknowledgments}
We thank Patrick Fox and Shunsaku Horiuchi for helpful conversations. I.H. is supported by generous contributions from Philip Rice and the Brinson Foundation. D.H. and G.K. are supported by the Fermi Research Alliance, LLC under Contract No.~DE-AC02-07CH11359 with the U.S. Department of Energy, Office of Science, Office of High Energy Physics. D.S. is supported by the Visiting Scholars Award Program of the Universities Research Association, the U.S. Department of Energy Office of Science under award number DE-SC0020262 and the Grants-in-Aid for Scientific Research (KAKENHI) No.~20H05852.
\end{acknowledgments}

\bibliography{DodelsonWidrowTwinHiggs}

\onecolumngrid
\appendix

\section{Derivation of the evolution equations} \label{derivation_appendix}

To determine how the temperature of the SM bath is impacted by a particle decaying out of equilibrium, we begin with the treatment presented in Sec.~5.3 of Ref.~\cite{KolbTurnerTheEarlyUniverse}. The second law of thermodynamics ($dS = dQ/T$) allows us to relate the change in total entropy, $S = s a^3$, in each sector to the heat $Q$ that is injected by decays of $N_A$. The first law of thermodynamics ($dQ = dE + P dV$) also lets us relate $Q$ to the change in internal energy, $E = \rho_{N_A}a^3$, and pressure $P$ of $N_A$, where $dV = d(a^3)$ is the change in volume due to the expansion of the Universe. The time derivative of the total entropy is then
\begin{equation} \label{dSdt_1}
    \begin{split}
    \frac{dS}{dt} = \frac{1}{T}\frac{dQ}{dt} &= -\frac{1}{T} \left( \frac{d(a^3 \rho_{N_A})}{dt} + P_{N_A} \frac{d(a^3)}{dt} \right) \\
    &= -\frac{1}{T} \left( \frac{d \rho_{N_A}}{dt} + 3H (\rho_{N_A} + P_{N_A}) \right) a^3 \\
    &= \frac{\angl{\Gamma_{N_A}} \rho_{N_A}}{T} a^3.
    \end{split}
\end{equation}
In the last equality, we simplify the expression by applying Eq.~(\ref{NA_equation}), leaving only a term dependent on the decay of $N_A$. We can also use the relationship between entropy density and temperature,
\begin{equation} \label{entropy_density}
    s(T) = \frac{2\pi^2}{45} g_{\star,s} T^3,
\end{equation}
to express the time derivative of the total entropy in another way:
\begin{equation} \label{dSdt_2}
    \begin{split}
        \frac{dS}{dt} = \frac{d(sa^3)}{dt} &= 3\dot{a}a^2 s + a^3 \dot{s}\\
        &= 3\dot{a}a^2 s + a^3 \frac{2\pi^2}{45}\left( 3\dot{T}T^2 g_{\star,s} + \dot{g}_{\star,s} T^3 \right)\\
        &= 3s a^3 \left( \frac{1}{a}\frac{da}{dt} + \frac{1}{T}\frac{dT}{dt} + \frac{1}{3g_{\star,s}} \frac{d g_{\star,s}}{dt} \right)\\
        &= 3Hs a^3 \left( 1 + \frac{a}{T}\frac{dT}{da} + \frac{1}{3}\frac{T}{g_{\star,s}}\frac{dg_{\star,s}}{dT} \frac{a}{T} \frac{dT}{da} \right)\\
        &= 3Hs a^3 \left( 1 + \frac{d\ln T}{d\ln a}\left[ 1 + \frac{1}{3} \frac{d\ln g_{\star,s}}{d\ln T} \right] \right).
    \end{split}
\end{equation}
By equating Eqs.~(\ref{dSdt_1}) and (\ref{dSdt_2}), we obtain a differential equation describing the temperature evolution of either sector with $a$:
\begin{equation} \label{temperature_evolution}
    \frac{d\ln T}{d\ln a} = -\frac{1 - \cfrac{\angl{\Gamma_{N_A}} \rho_{N_A}}{3HsT}}{1 + \cfrac{1}{3} \cfrac{d\ln g_{\star,s}}{d\ln T}}.
\end{equation}

In the limit of $\rho_{N_A} = 0$ and constant $g_{\star,s}$, Eq.~(\ref{temperature_evolution}) becomes $-1$, corresponding to the standard relationship for radiation, $T \propto a^{-1}$. Note that this derivation is valid only after the SM and twin sectors have decoupled.

\end{document}